\documentclass[12pt]{article}

\usepackage[english]{babel}
\usepackage{graphicx}   

\title{The dog-and-rabbit chase problem as 
an exercise in introductory kinematics}

\author{ O.~I.~Chashchina \\ 
Department of physics, Novosibirsk State University \\ 630 090,
Novosibirsk, Russia \vspace*{4mm} \\
Z.~K.~Silagadze \\
Budker Institute of Nuclear Physics and \\
Novosibirsk State University, 630 090, Novosibirsk, Russia }

\date{}

\begin{document}

\maketitle

\begin{abstract}
The purpose of this article is to present a simple solution of the classic 
dog-and-rabbit chase problem which emphasizes the use of concepts of 
elementary kinematics and, therefore, can be used in introductory mechanics 
course. The article is based on the teaching experience of introductory 
mechanics course at Novosibirsk State University for first year physics 
students which are just beginning to use advanced mathematical methods in 
physics problems. We hope it will be also useful for students and teachers 
at other universities too.

\end{abstract}

\section{Introduction}
\begin{figure}[htb]
\centerline{\includegraphics[height=75mm]{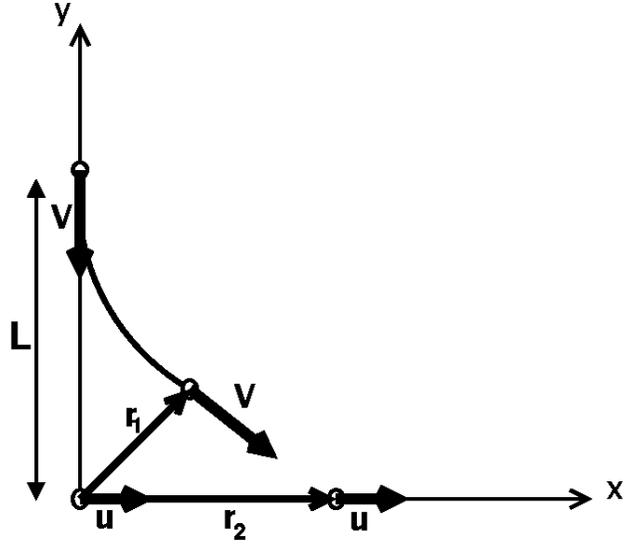}}
\caption{Dog and rabbit chase. The dog is heading always towards
the rabbit.}
\label{fig1}
\end{figure}

A rabbit runs in a straight line with a speed $u$. A dog with a speed
$V>u$ starts to pursuit it and during the pursuit always runs in the
direction towards the rabbit. Initially the rabbit is at the origin while
the dog's coordinates are $x(0)=0,\;y(0)=L$ (see Fig.\ref{fig1}). After
what time does the dog catch the rabbit?

This classic chase problem and its variations are often used in introductory 
mechanics course \cite{Irodov,Belchenko,Silagadze}. When one asks to find
the dog's trajectory (curve of pursuit), the problem becomes an exercise in
advanced calculus and/or in the elementary theory of differential equations
\cite{Sbornik,Ptak}. However, its treatment simplifies if the traditional
machinery of physical kinematics is used \cite{Mungan}.

The mathematics of the solution becomes even simpler if we further underline
the use of physical concepts like reference frames, vector equations, 
decomposition of velocity into radial and tangential components.

\section{Duration of the chase}
Let $\vec{r}_1$ be a radius-vector of the dog and  $\vec{r}_2$ -- a 
radius-vector of the rabbit. So that 
\begin{equation}
\dot {\vec{r}}_1=\vec{V},\;\;\;\dot {\vec{r}}_2=\vec{u}. 
\label{eq1}
\end{equation}
As the dog always is heading towards the rabbit, we can write
\begin{equation}
\vec{r}_2-\vec{r}_1=k(t)\vec{V}.
\label{eq2}
\end{equation}
The proportionality coefficient $k(t)$ depends on time. Namely, at the start 
and at the end of the chase we, obviously, have
\begin{equation}
k(0)=\frac{L}{V},\;\;\;k(T)=0.
\label{eq3}
\end{equation}
Differentiating (2) and using (1), we get
\begin{equation}
\vec{u}-\vec{V}=\dot k(t)\vec{V}+k(t)\dot{\vec{V}}.
\label{eq4}
\end{equation}
As the dog's velocity does not change in magnitude, it must be perpendicular 
to the dog's acceleration all the time:
\begin{equation}
\vec{V}\cdot\dot{\vec{V}}=0.
\label{eq5}
\end{equation}
Formally this follows from
$$0=\frac{dV^2}{dt}=\frac{d\vec{V}^2}{dt}=2\vec{V}\cdot\dot{\vec{V}}.$$
Equations (4) and (5) imply
$$\vec{V}\cdot\left (\vec{u}-\vec{V}\right )=\dot k(t) \vec{V}\cdot  
\vec{V},$$
or
\begin{equation}
uV_x-V^2=\dot k(t)V^2.
\label{eq6}
\end{equation}
We can integrate the last equation and get
\begin{equation}
ux-V^2t=k(t)V^2-LV.
\label{eq7}
\end{equation}
At the time $t=T$, when the dog catches the rabbit and the chase terminates,
we must have $x=uT$, and remembering that $k(T)=0$ we easily find $T$ from
(7):
\begin{equation}
T=\frac{LV}{V^2-u^2}.
\label{eq8}
\end{equation}

\section{Dog's trajectory in the rabbit's frame}
\begin{figure}[htb]
\centerline{\includegraphics[height=50mm]{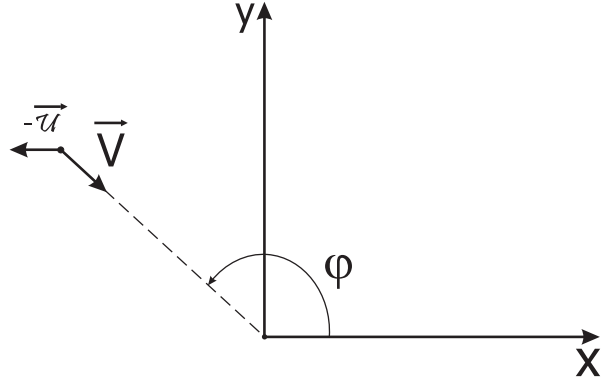}}
\caption{Components of the dog's velocity in the rabbit's frame}
\label{fig2}
\end{figure}

Let us decompose the dog's velocity into the radial and tangential components 
in the rabbit's frame (see Fig.\ref{fig2}): $V_r=-V+u\cos{(\pi-\varphi)}=
-V-u\cos{\varphi}$, $ V_\varphi=u\sin{(\pi-\varphi)}=u\sin{\varphi}$.
But $V_r=\dot{r}$ and $V_\varphi=r\dot{\varphi}$. Therefore
\begin{equation}
\dot{r}=-V-u\cos{\varphi},\;\;\;\;\;\; r\dot{\varphi}=u\sin{\varphi}.
\label{eq9}
\end{equation}
If we divide the first equation on the second and take into account that
$$\frac{\dot{r}}{\dot{\varphi}}=\frac{dr}{d\varphi},$$
we get
$$\frac{1}{r}\frac{dr}{d\varphi}=-\frac{V+u\cos{\varphi}}{u\sin{\varphi}}.$$
Hence
$$\ln{\frac{r}{L}}=-\int\limits_{\pi/2}^\varphi \frac{V+u\cos{\varphi}}
{u\sin{\varphi}}\,d\varphi.$$
Putting $z=\cos{\varphi}$ and using the decomposition
$$\frac{\frac{V}{u}+z}{1-z^2}=\frac{A}{1-z}+\frac{B}{1+z},$$
where
$$A=\frac{1}{2}\left (1+\frac{V}{u}\right ),\;\;
B=\frac{1}{2}\left (\frac{V}{u}-1 \right ),$$
we can easily integrate and get
$$
-\int\limits_{\pi/2}^\varphi \frac{V+u\cos{\varphi}}{u\sin{\varphi}}\,
d\varphi=\ln{\frac{(1+\cos{\varphi})^B}{(1-\cos{\varphi})^A}}= $$ $$  
\ln{\left [ \left (\cot{\frac{\varphi}{2}}\right )
^\frac{V}{u}\frac{1}{\sin{\varphi}}\right ]}. 
$$
Therefore, the equation of the dog's trajectory in the rabbit's frame is
\begin{equation}
r=\frac{L}{\sin{\varphi}}\left (\cot{\frac{\varphi}{2}}\right )
^\frac{V}{u}.
\label{eq10}
\end{equation}

\section{Dog's curve of pursuit} 
Let us now find the dog's trajectory in the laboratory frame. From (9)
and (10), we get 
$$\frac{L\dot{\varphi}}{\sin{\varphi}}\left (\cot{\frac{\varphi}{2}}\right )
^\frac{V}{u}=u\sin{\varphi},$$
which, by using
$$\frac{d\varphi}{\sin^2{\varphi}}=-\frac{1}{2}\left [ 1+\frac{1}
{\cot^2{\frac{\varphi}{2}}}\right]\,d\left(\cot{\frac{\varphi}{2}}\right), $$
can be recast in the form
$$\left [ 1+\frac{1}{\cot^2{\frac{\varphi}{2}}}\right]
\left (\cot{\frac{\varphi}{2}}\right )^\frac{V}{u}\,d\left(\cot{\frac{\varphi}
{2}}\right)=-\frac{2u}{L}\,dt.$$
Therefore
\begin{equation}
\frac{t}{T}=1-\frac{1}{2\nu}\left [ (\nu-1)\left (\cot{\frac{\varphi}{2}}
\right )^{\nu+1}+(\nu+1)\left (\cot{\frac{\varphi}{2}}\right )^{\nu-1}\right ],
\label{eq11}
\end{equation}
where
$$\nu=\frac{V}{u},$$
and $T$ was defined earlier through (8).
But by (10)
$$y=r\sin{\varphi}=L\left ( \cot{\frac{\varphi}{2}}\right )^\nu, $$
and, therefore, (11) reproduces the result of Ref.~\cite{Mungan}
\begin{equation}
\frac{t}{T}=1-\frac{1}{2}\left [ (1-\tilde\nu)\left (\frac{y}{L}
\right )^{\tilde\nu+1}+(1+\tilde\nu)\left (\frac{y}{L}\right )^{1-\tilde\nu}
\right ],
\label{eq12}
\end{equation}
where
$$\tilde\nu=\frac{1}{\nu}=\frac{u}{V}.$$
From (7), we have
\begin{equation}
\frac{x}{L}=\frac{\nu^3}{\nu^2-1}\left (\frac{t}{T}+\frac{k(t)}{T}\right )
-\nu
\label{eq13}
\end{equation}
and to get the equation of the dog's trajectory, we have to express $k(t)$ 
through $y$. Taking the $y$-component of the vector equation (2), we 
get
$$-y=k(t)\,\dot{y},$$
and therefore
\begin{equation}
k(t)=-\frac{y}{\dot{y}}.
\label{eq14}
\end{equation}
It remains to differentiate (12) to get $\dot{y}$, and hence the 
desired expression for $k(t)$:
\begin{equation}
\frac{k(t)}{T}=-\frac{1}{2}\left (\tilde{\nu}^2-1\right)\left [ \left(
\frac{y}{L}\right)^{1+\tilde\nu}+\left(\frac{y}{L}\right)^{1-\tilde\nu}
\right ].
\label{eq15}
\end{equation}
Substituting (12) and (15) into (13), we finally get
the dog's curve of pursuit in the form
\begin{equation}
\frac{x}{L}=\frac{\tilde\nu}{1-\tilde{\nu}^2}+\frac{1}{2}\left [
\frac{1}{1+\tilde\nu}\left (\frac{y}{L}\right )^{1+\tilde\nu}-
\frac{1}{1-\tilde\nu}\left (\frac{y}{L}\right )^{1-\tilde\nu}\right ].
\label{eq16}
\end{equation}
Of course, this result is the same as found earlier in the literature 
\cite{Sbornik,Ptak,Mungan}, up to applied conventions.

\section{The limit distance for equal velocities}
\begin{figure}[htb]
\centerline{\includegraphics[height=60mm]{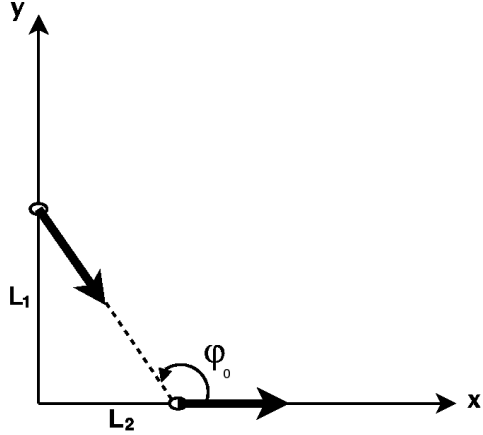}}
\caption{The chase with equal velocities}
\label{fig3}
\end{figure}

Let speeds of the dog and the rabbit are equal in magnitude and their initial 
positions are as shown in the Fig.\ref{fig3}. To what limit converges the 
distance between them?

It is convenient to answer this question in the rabbit's frame. For equal 
velocities $V=u$, equations (9) take the form
$$\dot{r}=-u(1+\cos{\varphi}),\; r\dot{\varphi}=u\sin{\varphi}.$$
Therefore
$$\dot{r}(1-\cos{\varphi})+r\dot{\varphi}\sin{\varphi}=-u(1-\cos^2{\varphi})+
u\sin^2{\varphi}=0.$$
But
$$\dot{r}(1-\cos{\varphi})+r\dot{\varphi}\sin{\varphi}=
\frac{d}{dt}(r-r\cos{\varphi}).$$
We see that
$$r(1-\cos{\varphi})=C,$$
where $C$ is a constant. At $t=0$ we have (see Fig.\ref{fig3})
$$r_0=\sqrt{L_1^2+L_2^2},\;\;\; \cos{\varphi_0}=-\frac{L_2}
{\sqrt{L_1^2+L_2^2}}.$$
Therefore 
$$C=L_2+\sqrt{L_1^2+L_2^2}.$$
In the rabbit's frame $\varphi\to\pi$, when $t\to\infty$. Hence the limit 
distance between the dog and the rabbit is \cite{Ptak}
$$r_{min}=\frac{C}{1-\cos{\pi}}=\frac{L_2+\sqrt{L_1^2+L_2^2}}{2}.$$

\section{Chase along the tractrix}
If the rabbit is allowed to change the magnitude of his speed, he can manage 
to keep the distance between him and the dog constant. Let us find the 
required functional dependence $u(t)$ and the dog's trajectory in this case.
If $r=L=const$, so that $\dot r=0$, equations (9) take the form
\begin{equation}
V=-u\cos{\varphi},\;\;\;L\,\dot\varphi=u\sin{\varphi}.
\label{eq17}
\end{equation}
Therefore
$$\frac{L}{V}\,\dot\varphi=-\tan{\varphi},$$
which can be easily integrated to get
$$\ln{\left (\frac{\sin{\varphi}}{\sin{\varphi_0}}\right )}=-\frac{V}{L}t,$$
or
$$\sin{\varphi}=\sin{\varphi_0}\;e^{-\frac{V}{L}t}.$$
Then the first equation of (17) determines the required form of the
rabbit's velocity
\begin{equation}
u(t)=\frac{V}{\sqrt{1-\sin^2{\varphi_0}\;e^{-\frac{2V}{L}t}}}
\label{eq18}
\end{equation}
(note that we are assuming $\varphi\ge\varphi_0>\pi/2$, as in Fig.\ref{fig3},
so that $\cos{\varphi}=-\sqrt{1-\sin^2{\varphi}}$).

Now let us find the dog's trajectory. We have (see Fig.\ref{fig3})
$$
x=\int\limits_0^t u(\tau)\,d\tau-L\cos{(\pi-\varphi)}=\int\limits_0^t u(\tau)
d\tau+L\cos{\varphi}, $$ $$
y=L\sin{(\pi-\varphi)}=L\sin{\varphi}. \nonumber
$$
But
\begin{equation}
\int\limits_0^t u(\tau)\,d\tau=L\int\limits_{\varphi_0}^\varphi\frac{d\varphi}
{\sin{\varphi}}=-L\int\limits_{\varphi_0}^\varphi\frac{d\cos{\varphi}}
{1-\cos^2{\varphi}},
\label{eq19}
\end{equation}
because
$$u(t)\,dt=\frac{L}{\sin{\varphi}}\,d\varphi,$$
according to the second equation of (17).

\noindent By using the decomposition
$$\frac{1}{1-\cos^2{\varphi}}=\frac{1}{2}\left [\frac{1}{1+\cos{\varphi}}+
\frac{1}{1-\cos{\varphi}}\right ],$$
the integral in (19) is easily evaluated with the result
$$\int\limits_0^t u(\tau)\,d\tau=L\left [\ln{\left (\cot{\frac{\varphi_0}{2}}
\right )}-\ln{\left (\cot{\frac{\varphi}{2}}\right )}\right ].$$
Therefore the parametric form of the dog's trajectory is
$$\frac{x}{L}=\cos{\varphi}+\ln{\left (\cot{\frac{\varphi_0}{2}}\right )}-
\ln{\left (\cot{\frac{\varphi}{2}}\right )},\;\;\;\frac{y}{L}=\sin{\varphi}.$$
To get the explicit form of the trajectory, we use
$$
\cos{\varphi}=-\sqrt{1-\frac{y^2}{L^2}}, $$ $$
\cot{\frac{\varphi}{2}}=
\frac{1+\cos{\varphi}}{\sin{\varphi}}=\frac{L-\sqrt{L^2-y^2}}{y}, \nonumber
$$
which gives
\begin{equation}
x=L\ln{\left (\cot{\frac{\varphi_0}{2}}\right )}-L\ln{\frac{L-\sqrt{L^2-y^2}}
{y}}-\sqrt{L^2-y^2}.
\label{eq20}
\end{equation}
This trajectory is a part of tractrix -- the famous curve \cite{Yates,Cady}
with the defining property that the length of its tangent, between its
directrix (the $x$-axis in our case) and the point of tangency, has the same 
value $L$ for all points of the tractrix.

\section{Concluding remarks}
We think the problem considered is of pedagogical value for undergraduate 
students, which take their first year course in physics. It demonstrates the 
use of some important concepts of physical kinematics, as already stressed 
by Mungan in Ref.~\cite{Mungan}. The approach presented in this article 
requires only minimal mathematical background and, therefore, is suitable for
students which just begin their physics education. However, if desired, this 
classic chase problem allows a demonstration of more elaborate mathematical  
concepts like Frenet-Serret formulas \cite{Puckette}, Mercator projection in 
cartography \cite{Pijls}, and even hyperbolic geometry (which is realized on 
the surface of revolution of a tractrix about its directrix) \cite{Bertotti}.
Interested reader can find some other variations of this chase problem 
in \cite{Lotka,Lalan,Lotka1,XX,Wunderlich}.

\section*{Acknowledgments}
The work is supported in part by grants Sci.School-905.2006.2 and
RFBR 06-02-16192-a.

\section*{Note added after publication}
Another simple and transparent method how to calculate the duration of the 
chase is given in \cite{Tarantsev}.

\end{document}